# Transmission Line Impedance of Carbon Nanotube Thin Films for Chemical Sensing

G. Esen and M. S. Fuhrer[a]
*Department of Physics and Center for Superconductivity Research, University of Maryland, College Park, MD 20742*

M. Ishigami and E. D. Williams
*Department of Physics and Materials Research Science and Engineering Center, University of Maryland, College Park, MD 20742*

We measure the resistance and frequency-dependent gate capacitance of carbon nanotube (CNT) thin films in ambient, vacuum, and under low-pressure ($10^{-6}$ torr) analyte environments. We model the CNT film as an RC transmission line and show that changes in the measured capacitance as a function of gate bias and analyte pressure are consistent with changes in the transmission line impedance due to changes in the CNT film resistivity alone; the electrostatic gate capacitance of the CNT film does not depend on gate voltage or chemical analyte adsorption. However, the CNT film resistance is enormously sensitive to low pressure analyte exposure.

a) Author to whom correspondence should be addressed. Electronic mail:

mfuhrer@umd.edu

Single-walled carbon nanotubes (CNTs) have a number of favorable properties for chemical sensors[1-5]: CNTs have high surface-to-volume ratio, with all atoms located on the surface; the extremely high carrier mobility[6] in semiconducting CNTs makes them excellent detectors of charge[7,8]; and the nanometer size of CNTs may allow ultra-thin sensitizing or concentrating layers, reducing diffusion times. Thin films of CNTs may be easily and inexpensively prepared on a variety of substrates, and CNT thin-film transistors (TFTs) can exhibit effective charge carrier mobilities comparable to single-crystal silicon[2]. While the presence of metallic CNTs in as-prepared CNT-TFTs limits their on-off ratio and application to logic circuits, such CNT-TFTs may still be suitable for chemical sensing applications.

Chemical sensing in a CNT film may take place through a number of different mechanisms. Adsorption of chemical analytes on or near the CNT film may change the charge density (through charge transfer, or "doping"), charge carrier mobility (through scattering), CNT-electrode contact resistance, CNT-CNT contact resistance, and/or gate capacitance. Sorting out these mechanisms is not trivial; for instance, calculation of the mobility requires knowledge of the gate capacitance, so changes in mobility and gate capacitance are not easily discriminated through electrical transport measurements alone.

Here we measure the changes in DC electrical characteristics of a CNT film upon exposure to chemical analytes, and simultaneously measure the gate capacitance of the CNT film directly at frequencies of 50 Hz-20 kHz using a sensitive capacitance bridge. We find that, due to the high impedance of the CNT film, the capacitance must be analyzed within the framework of an RC transmission line consisting of the resistive CNT film capacitively coupled to the gate electrode. Within this framework, we can decouple the resistive and capacitive response, and we show that changes in the measured capacitance as a function of gate bias and analyte pressure are consistent with changes in the capacitive part of the transmission line impedance due to changes in the CNT film resistivity alone; the electrostatic gate capacitance of the CNT film does not depend on gate voltage or chemical analyte adsorption to within the resolution of our measurement. However, the resistance of the CNT film is enormously sensitive to exposure to small pressures ($< 10^{-6}$ Torr) of analytes, and gate-voltage-dependent resistivity changes show analyte-dependent signatures.

CNT films were grown by chemical vapor deposition[9,10] on $t = 500$ nm thick $SiO_2$ on degenerately doped Si substrates which serve as the back gate. Following growth, devices were fabricated using two steps of photolithography; the first formed the interdigitated source and drain electrodes (5 nm Cr and 55 nm Au), the second masked the device region with photoresist for removal of excess CNTs via oxygen plasma. Figure 1 (a) shows a completed device and Figure 1 (b) shows the CNT film morphology. The sample was first measured in ambient, then placed into an ultra high vacuum (UHV) chamber equipped with a mass spectrometer to calibrate the partial pressure of the analytes. The sample was out-gassed by heating above 200 $C^o$ in UHV and then cooling to room temperature before and after each gas exposure. Conductance was measured by applying a DC voltage and measuring current with an Ithaco 1211 transimpedance amplifier. Capacitance between source and gate was measured using an Andeen Hagerling 2700A capacitance bridge, with drain electrode floating.

The CNT thin film is composed of individual CNTs having a resistance which, in the case of semiconducting nanotubes, depends on the gate voltage, and connected by

CNT-CNT junctions[11]. While such a network has discrete elements, over the gate length of our device (500 μm) it is reasonable to treat the CNT network as a continuous resistive film with resistance per length $r = R/L$, back-gate electrostatic capacitance per length $c_{el} = C_{el}/L = \kappa\varepsilon_0 w/t$, and inductance per length $l = \mu_0 t$, where κ is the relative permittivity (3.9 for SiO$_2$), $\varepsilon_0$ the electric constant, $\mu_0$ the magnetic constant, and $w$ the device width (≈8 mm). At low frequency $\omega \ll r/l$, such a network may be modeled as an RC transmission line: a signal at frequency ω decays along the line with a characteristic decay length

$$l_0 = \sqrt{\frac{2}{r\omega c_{el}}}.$$

For $L \gg l_0$, the line has a complex impedance to the gate

$$Z = \frac{(1-i)}{\sqrt{2}}\sqrt{\frac{r}{\omega c_{el}}}.$$

Comparing $Z$ to a series RC circuit, that is $Z = R_{CNT} - \frac{i}{\omega C_{CNT}}$, the effective capacitance of the film is $C_{CNT} = \sqrt{\frac{2c_{el}}{\omega r}} = c_{el} l_0$. Note that $C_{CNT}$ may be thought of as the gate capacitance of a section of the transmission line equal to the decay length $l_0$.

The main panel of Figure 1 (c) shows the measured capacitance $C$ as a function of frequency from 50 Hz to 20 kHz for a typical device in ambient atmosphere at various gate voltages $V_g$. The DC device resistance per length $r$ varies monotonically from 7.6x10$^7$ to 1x10$^9$ Ω/m for $V_g$ = -10 V and +10 V respectively. We explain the behavior as follows. At low frequency, the measured gate capacitance is the sum of the source and drain electrode capacitances $C_s$ and $C_d$, and the capacitance of the entire CNT film, $C_{CNT}$. As the frequency increases, the characteristic length of the RC transmission line $l_0$ decreases below the length of the device $L$, the measured capacitance becomes $C_s + C_{CNT}$, with $C_{NT} = \sqrt{\dfrac{2c}{\omega r}} \propto \omega^{-1/2}$. The measured capacitance indeed drops as $\omega^{-1/2}$ at high frequency. The crossover is expected at approximately $l_0 = L$, or $\omega/2\pi = 3.5 - 50$ kHz, which is reasonably close to the observed crossover.

We can see from Figure 1 (c) that at high frequencies the capacitance depends not only on frequency, but on gate voltage, becoming lower when the gate voltage is more positive (off state). This has been interpreted[12] as due to the quantum capacitance of the semiconducting CNT becoming small in the off state, decreasing the total capacitance. However, the RC transmission line model predicts that $C_{CNT} \propto r^{-1/2}$. Figure 2 plots both $C_{CNT}(14 \text{ kHz})$ and $R(DC)^{-1/2}$ vs. $V_g$. We see that $C_{CNT}(V_g)$ follows the variation of $R(V_g)$, implying that the electrostatic capacitance of the CNT film is *independent* of $V_g$. This is consistent with the lack of dependence of $C$ on $V_g$ at low frequency in Fig. 1 (c). The absence of a large quantum capacitance effect is puzzling, but likely implies that there is always a significant density of states at the Fermi level in the semiconducting CNTs at all $V_g$.

We now turn to measurements in UHV and small analyte gas pressures. Figures 3(a) and 3(b) show the effect of acetone on $C_{CNT}(20 \text{ kHz})$ and $R(DC)$ respectively of the CNT network as a function of gate voltage. Measurements were taken approximately 2-3 minutes after introduction of the analyte; $C_{CNT}(20 \text{ kHz})$ and $R(DC)$ were not observed to vary with time. We find in general that $R$ in UHV is significantly higher than in ambient (see inset, Figure 2), and $R(V_g)$ shows more symmetric ambipolar behavior. (Changes in $C_{CNT}(V_g)$ from ambient to UHV follow the variation of $R(V_g)^{-1/2}$ as above.) It is not immediately clear from Figures 3(a) and 3(b) that changes in $C_{CNT}$ and $R$ upon acetone exposure are quantitatively related. However, we may again look at the predictions of the

RC transmission line model. Since $C_{CNT} = \sqrt{\frac{2c_{el}}{\omega r}}$, we expect for small changes that

$$\frac{\Delta C}{C} = -\frac{1}{2}\frac{\Delta R}{R},$$ where $\Delta C = C_{CNT}(\text{analyte}) - C_{CNT}(\text{UHV})$ and

$\Delta R = R(\text{analyte}) - R(\text{UHV})$ [13]. In figure 4(a) we plot $\frac{\Delta C}{C}$ and $-\frac{1}{2}\frac{\Delta R}{R}$ due to acetone introduction in the chamber as a function of gate voltage. The vertical axis in the figure 4(a) is showing the changes of both capacitance and resistance. Note that the resistance is measured at DC, and therefore is a true probe of the film resistivity $r$, and is not altered by the RC transmission line impedance. From figure 4(a) we conclude that measured capacitance change of the CNT network upon acetone exposure may be entirely explained by the change in resistivity of the film and concomitant change in the RC transmission line impedance.

Similar to the case of acetone, in figure 4(b) and 4(c) we plot $\frac{\Delta C}{C}$ and $-\frac{1}{2}\frac{\Delta R}{R}$ as a function of gate voltage due to water introduction at $2 \times 10^{-8}$ Torr and argon introduction at $3 \times 10^{-6}$ Torr, respectively. Figure 4(b) indicates that the capacitance and resistance changes are again related to each other, and the capacitance change is due to the resistivity change of the CNT network. Figure 4(c) shows that introduction of argon (a nonpolar and noble gas which is not expected to have any effect on either the scattering or the polarization) has no effect on capacitance or resistance. Thus figure 4(c) may be used as a baseline response with which to compare figures 4(a) and 4(b).

It is notable that the resistance of the CNT networks is enormously sensitive to exposure to low pressures (< $10^{-6}$ Torr) of various analytes, and the gate voltage dependence of the resistance change is different for different analytes, indicating a possible means for chemical selectivity. The resistance change upon acetone introduction is presumably due to increased scattering due to weakly adsorbed molecules on the CNT surface, or at the CNT-CNT junctions . At low concentrations, we expect the former is the most likely, due to the much larger area of CNT compared to the junctions. This offers the immediate possibility of quantifying the adsorbate concentration by measuring temperature and pressure dependence of the resistivity changes, and hence studying the magnitude and carrier-density dependence of scattering due to adsorbates on CNT; such experiments are underway.

This research was supported by the Laboratory for Physical Sciences, the U.S. Army Research Laboratory MICRA Program, and the NSF-MRSEC Shared Equipment Facilities. MI received support from the Director of Central Intelligence Postdoctoral Fellowship program. We thank Alma Wickenden and Neil Goldsman for useful discussions and Jian Hao Chen for help with the experiments.

Figure Captions

**Figure 1**. (a) Scanning electron microscope (SEM) image of the CNT device. Scale bar is 1 mm. (b) SEM image of as-grown CNT film. Scale bar is 20 microns. (c) Device capacitance as a function of frequency at different gate voltages in ambient atmosphere. The RMS amplitude of applied AC voltage during the measurement was 0.1 Volts. The solid line has a slope of -1/2. Inset shows a schematic of the capacitances in the device.

**Figure 2**. Capacitance $C_{CNT}$ and inverse square root of resistance $R^{-1/2}$ of the CNT film as a function of gate voltage $V_g$. Both data are taken in UHV (< $10^{-10}$ Torr pressure). $C_{CNT}$ is measured by applying 14 kHz sinusoidal drive voltage at 0.1 V RMS, and R by applying 500 mV source-drain bias. A constant 195 pF (approximately equal to the capacitance of the contacts) has been subtracted from the measured capacitance to obtain $C_{CNT}$. Inset shows $R(V_g)$. Black arrows show direction of $V_g$ sweeps.

**Figure 3**. Capacitance $C_{CNT}$ and resistance $R$ of the CNT film in UHV and on acetone exposure. (a) $C_{CNT}$ in UHV and 9.5 $10^{-7}$ Torr acetone pressure. (b) R in UHV and 9.5 x $10^{-7}$ Torr acetone pressure. $C_{CNT}$ data is taken at 20 KHz using 0.1 V drive voltage. Black arrows show direction of $V_g$ sweeps.

**Figure 4**. Capacitance and resistance changes of the CNT film device relative to UHV environment due to (a) 9.5 x $10^{-7}$ Torr acetone, (b) 2x$10^{-8}$ Torr water, and (c) 3x$10^{-6}$ Torr argon. Black arrows show direction of $V_g$ sweeps.

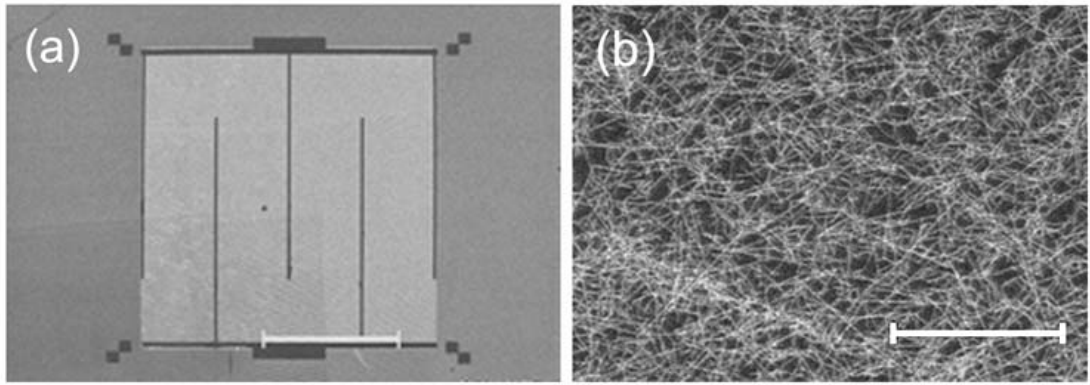

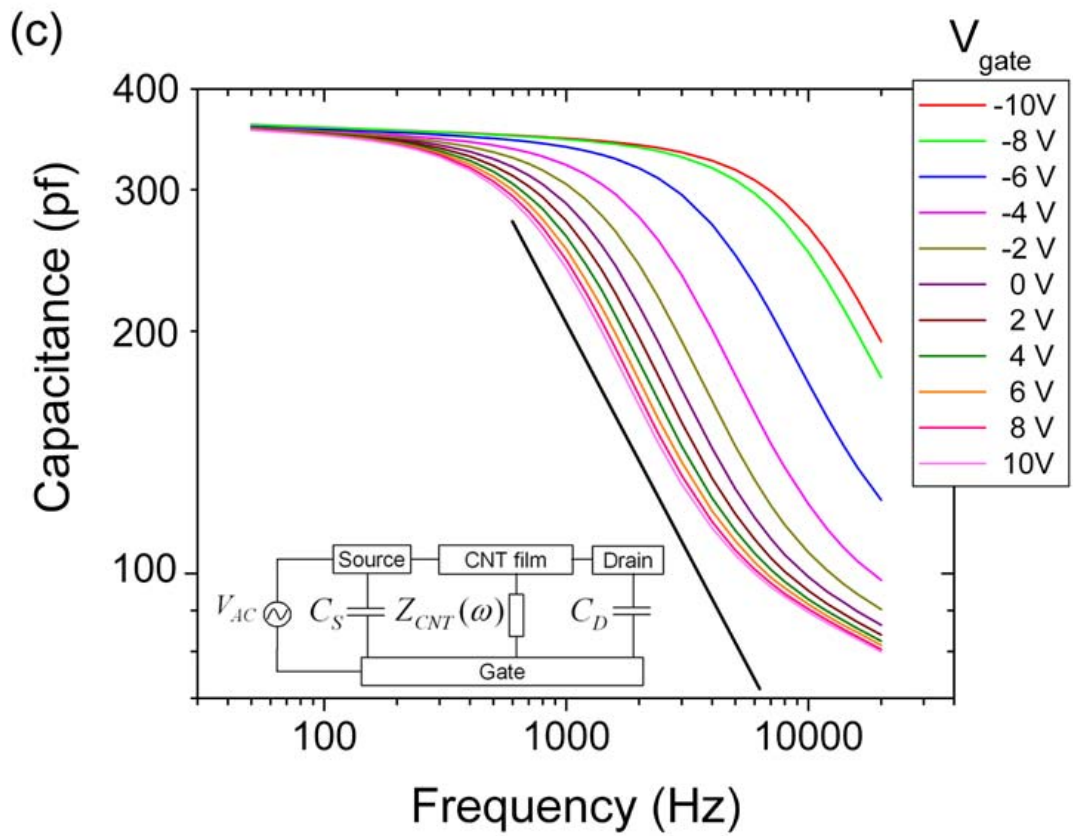

Figure 1.

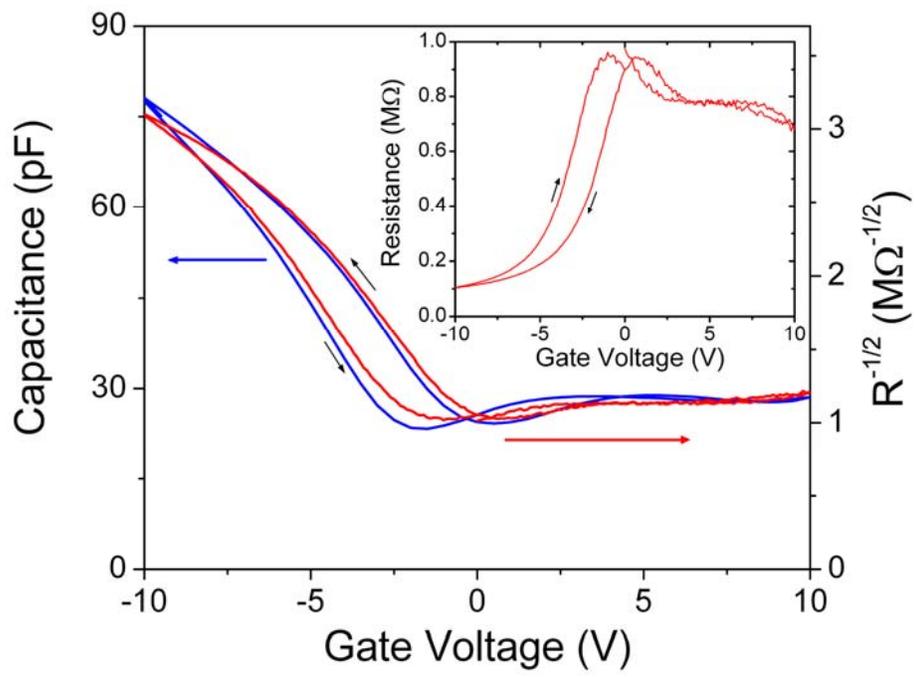

Figure 2.

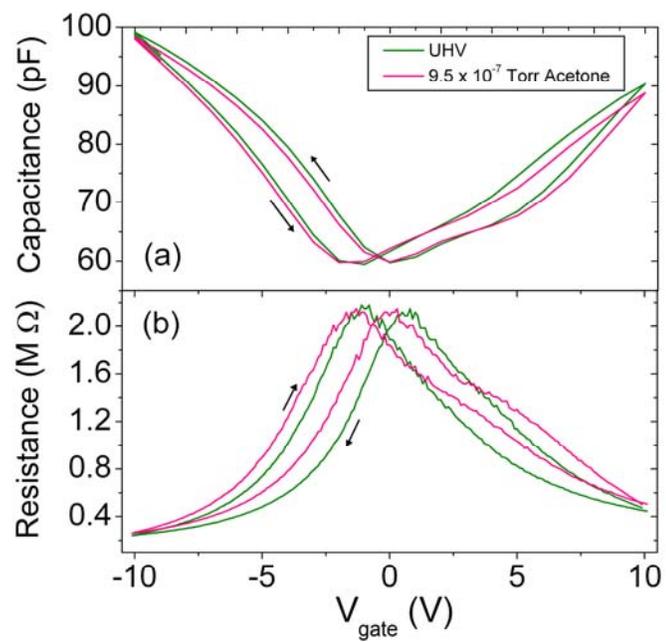

Figure 3.

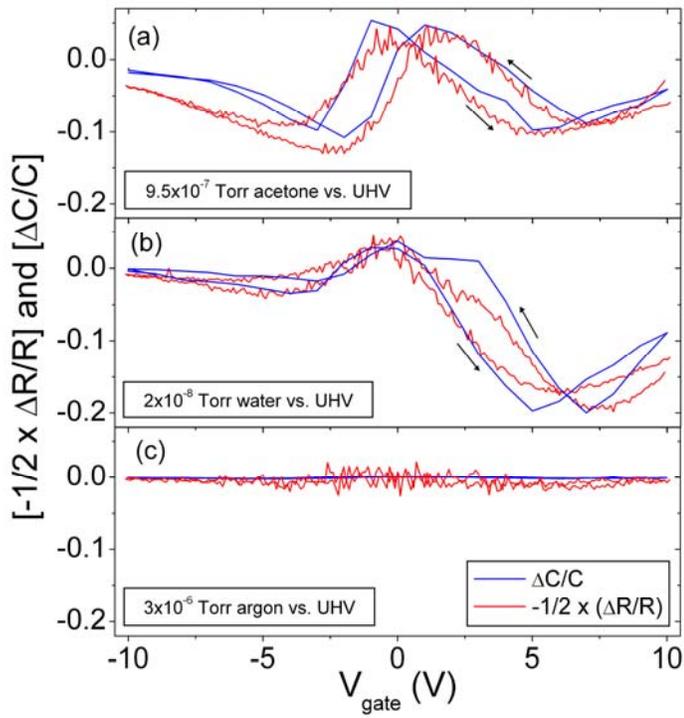

Figure 4.